\documentclass[aps,pra,reprint,floatfix,amsmath,amssymb,superscriptaddress,
]{revtex4-1}
\usepackage[english]{babel}
\usepackage{color}
\usepackage{times,amsthm}
\usepackage{graphicx}
\usepackage[caption=false]{subfig} 
\usepackage{braket}

\begin{document}
\title{Confined Contextuality in Neutron Interferometry: \\ Observing the Quantum Pigeonhole Effect}

\author{Mordecai Waegell}
\affiliation{Institute for Quantum Studies, Chapman University, Orange, CA 92866, USA}
\affiliation{Schmid College of Science and Technology, Chapman University, Orange, CA 92866, USA}

\author{Tobias Denkmayr}
\affiliation{Atominstitut, TU-Wien, Stadionallee 2, 1020 Vienna, Austria}

\author{Hermann Geppert}
\affiliation{Atominstitut, TU-Wien, Stadionallee 2, 1020 Vienna, Austria}

\author{David Ebner}
\affiliation{Atominstitut, TU-Wien, Stadionallee 2, 1020 Vienna, Austria}

\author{Tobias Jenke}
\affiliation{Institut Laue-Langevin 6, Rue Jules Horowitz, 38042 Grenoble Cedex 9, France}

\author{Yuji Hasegawa}
\affiliation{Atominstitut, TU-Wien, Stadionallee 2, 1020 Vienna, Austria}

\author{Stephan Sponar}
\affiliation{Atominstitut, TU-Wien, Stadionallee 2, 1020 Vienna, Austria}

\author{Justin Dressel}
\affiliation{Institute for Quantum Studies, Chapman University, Orange, CA 92866, USA}
\affiliation{Schmid College of Science and Technology, Chapman University, Orange, CA 92866, USA}

\author{Jeff Tollaksen}
\affiliation{Institute for Quantum Studies, Chapman University, Orange, CA 92866, USA}
\affiliation{Schmid College of Science and Technology, Chapman University, Orange, CA 92866, USA}



\begin{abstract}
Previous experimental tests of quantum contextuality based on the Bell-Kochen-Specker (BKS) theorem have demonstrated that not all observables among a given set can be assigned noncontextual eigenvalue predictions, but have never identified which specific observables must fail such assignment. We now remedy this shortcoming by showing that BKS contextuality can be confined to particular observables by pre- and postselection, resulting in anomalous weak values that we measure using modern neutron interferometry. We construct a confined contextuality witness from weak values, which we measure experimentally to obtain a $5\sigma$ average violation of the noncontextual bound, with one contributing term violating an independent bound by more than $99\sigma$. This weakly measured confined BKS contextuality also confirms the quantum pigeonhole effect, wherein eigenvalue assignments to contextual observables apparently violate the classical pigeonhole principle.
\end{abstract}

\maketitle

\section{Introduction}
Quantum contextuality, as introduced by Bell, Kochen and Specker (BKS) \cite{Bell2,KS}, forbids all observable properties of a system from being predefined independently from how they are observed. This phenomenon is one of the most counterintuitive aspects of quantum mechanics, and finds itself at the heart of recent quantum information processing applications \cite{howard2014contextuality, bechmann2000quantum,cabello2011hybrid,abbott2012strong, spekkens2009preparation,W_Primitive}.  
The BKS theorem is proved by exhibiting a \emph{BKS-set} of observables \cite{WA_3qubits,WA_Nqubits} that contains geometrically related and mutually commuting subsets (or measurement \emph{contexts}) that result in a logical incompatibility: Any noncontextual hidden variable theory (NCHVT) that pre-assigns eigenvalues globally to the entire BKS-set (i.e., \emph{non}contextually) results in a contradiction with the predictions of quantum mechanics. That is, at least one eigenvalue in a global assignment to a BKS-set cannot be predefined without violating a constraint on the product of eigenvalues within some context, which we call the \emph{contradictory context}.  See Appendix \ref{A:Theory} for a \ review of the BKS theorem and how the BKS-sets used in this article prove it.

\begin{figure}[b]
\includegraphics[width=\columnwidth]{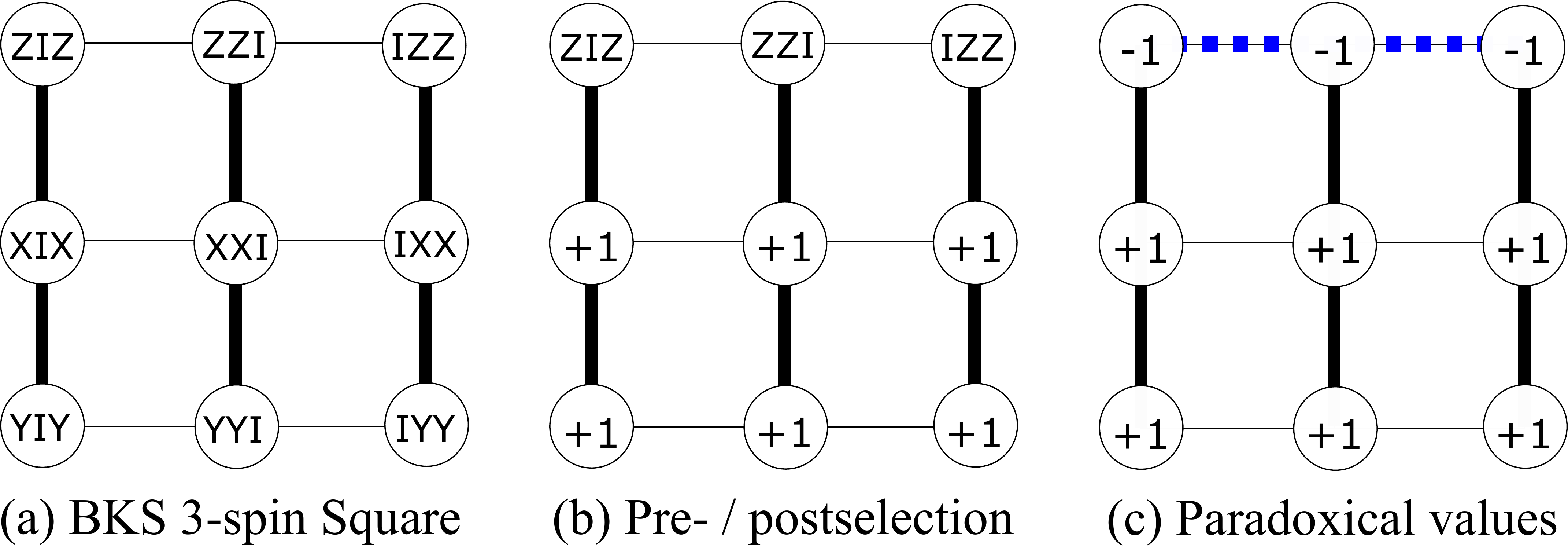}
\caption{Confining Bell-Kochen-Specker (BKS) contextuality in the 3-spin Square.  Each row or column (measurement context) of the Square mutually commutes. (a) According to quantum mechanics, the product of the three 3-spin measurement outcomes in each row is $+1$ (thin line), while their product in each column is $-1$ (thick line). (b) A particular preparation and postselection fixes the values of two rows. (c) In any noncontextual hidden variable theory, the remaining values must be $-1$, which confines the BKS contradiction to the top row (blue dashed line). This also demonstrates the quantum pigeonhole paradox: all pairs in the row appear anticorrelated, which violates the classical pigeonhole principle. Weak measurements confirm the paradox, revealing the correlation of each pair to be $-0.972 \pm 0.132$, $-1.050 \pm 0.140$, and $-1.020 \pm 0.137$, from left to right.}
\label{SquareSteps}
\end{figure}

Previous contextuality experiments \cite{kirchmair2009state,bartosik2009,d2013experimental} have confirmed such a global contradiction. However, neither the BKS theorem, nor these experiments, specify which contexts are contradictory.
In this article, using recently developed weak measurement techniques in neutron interferometry \cite{rauch00,rauch02,hasegawa03,klepp2014,denkmayr2014,sponar2015,denkmayr2016experimental}, we experimentally demonstrate 
which specific measurement context within a BKS-set (Fig.~\ref{SquareSteps}a) must contain contradictory value assignments, essentially \emph{confining} the contextuality \cite{waegell2015contextuality}. Like squeezing a balloon, we condition the BKS-set through pre- and postselection (Fig.~\ref{SquareSteps}b) to force the contradiction to appear in a particular context (Fig.~\ref{SquareSteps}c) \cite{cabello1997no,PhysRevLett.95.200405,pusey2015logical}. Remarkably, measuring the weak values \cite{aharonov1988result} within that context explicitly reveals the contradiction that is left implicit in the original BKS proof. The measured weak values violate the classical pigeonhole principle \cite{aharonov2016quantum}, and contradict NCHVT value assignments to the projectors in that context, which we call \emph{forbidden projectors}. We show that the confinement of contextuality in the quantum pigeonhole effect forces some of the forbidden projectors to have negative weak values. The appearance of these negative weak values thus witnesses the confined contextuality, making the forbidden projectors \emph{witness observables} for contextuality. These witnesses corroborate recent results \cite{pusey2014anomalous,mazurek2015experimental,Piacentini2016} that link negative projector weak values to contextuality using Spekkens' generalization of contextuality \cite{spekkens2005}, which encompasses the original notion of BKS. 

In our experiment, we witnessed the BKS-contextuality of neutron spin. We measured the spin using neutron interferometry by performing path-dependent spin rotations, making the path a weakly-coupled meter for the spin (Fig.~\ref{fig:setup}); conditioning the path measurements on spin postselections then reveals the desired weak values \cite{sponar2015}. We collected seventeen independent data sets of neutron spin measurements, indexed $n=1,\ldots,17$. We use these single-spin data sets to show confined contextuality within the $N$-spin Wheel BKS-sets \cite{WA_Nqubits} (see Appendix \ref{A:Theory}) for odd numbers of spins $N=3,5,\ldots,17$, using data sets $n=1,\ldots,N$ and the following simplification. While most contexts in an $N$-spin Wheel are entangled, we use separable pre- and postselection to fix the eigenvalues of certain observables (as in Fig.~\ref{SquareSteps}b), and by the definition of noncontextuality, any NCHVT must assign the same eigenvalues to those observables in the entangled contexts as it does in the separable contexts --- just as in \cite{cabello1997no}.  This would be true even if we were considering $N$ 2-level quantum systems of all different types --- photons, superconducting qubits, trapped ions, diamond NV-centers, and so on, regardless of where or when they are located with respect to one another --- it is a logical consequence of noncontextuality, which has nothing to do with actually performing joint measurements.  We weakly measure only the remaining separable observables, noting that the weak values are noncontextual by definition. As such, we are able to treat each of the single-spin data sets as representing a distinct spin within an $N$-spin Wheel BKS-set.  We are not claiming this is the same as performing genuine $N$-spin measurements; we are claiming that $N$ single-spin measurements are sufficient to reveal the contradiction inherent in the $N$-spin Wheel BKS set, between quantum mechanics and NCHVTs.

Weak values do not appear shot-by-shot in our experiment, but only as conditioned averages from ensembles of identically preselected and postselected data. The fact that we are constructing $N$-spin weak values from single-spin weak value measurements may seem odd, but this construction is generally valid for any averages from probability distributions describing independent (separable) systems --- including our distinct sets of neutron measurements.  That is, the complete set of collected single-neutron data was naturally divided into 17 smaller and independent subsets, each collected sequentially in time to minimize experimental drift, and each chosen to be sufficiently large to achieve acceptable statistical error for estimating a single-spin weak value. The number 17 was limited only by the total collected statistics, which was limited by the neutron flux from the reactor and the stability of the experimental setup. The measured $N$-spin witnesses violate their noncontextuality bounds by $\agt\!5\sigma$, showing that the contextuality was indeed confined in our experiment  (Fig.~\ref{WeakValues}). One particular 5-spin weak value exceeded its bound by more than $99 \sigma$. The $N$-spin confined contextuality observed here also violates the classical pigeonhole principle for putting $N$ pigeons in 2 boxes.  Our experiment verifies the quantum predictions, and rules out NCHVTs of quantum mechanics.

This article is organized as follows. In Section~\ref{sec:confine}, we discuss how to confine contextuality to particular contexts with pre- and postselection, and the relationship between such confined contextuality and the quantum pigeonhole effect. In Section~\ref{sec:witness} we show how the confined contextuality of the Wheel family of BKS-Sets permits the construction of $N$-spin contextuality witnesses that may be factored into measurable single-spin weak values. In Section~\ref{sec:procedure}, we detail the experimental procedure used to measure the single-spin weak values using neutron interferometry. In Section~\ref{sec:results}, we summarize the main results for the $N$-spin contextuality witnesses. We conclude in Section~\ref{sec:conclusion}. For completeness, we also include two appendices. In Appendix~\ref{A:Theory}, we describe the construction of the Wheel family of BKS-sets used in the main text. In Appendix~\ref{B:Experiment}, we provide additional details about the experimental determination of the single-spin weak values used to construct the results reported in the main text.

\section{Confining BKS contextuality}\label{sec:confine}%
BKS contextuality confinement follows from the Aharonov-Bergmann-Lebowitz (ABL) formula \cite{aharonov1964time}, which gives the probability of obtaining a particular strong measurement outcome $j$, between a preparation $\ket{\psi}$ and a postselection $\bra{\phi}$. The outcome $j$ corresponds to a projection operator $\Pi_j$ that is part of a complete measurement basis $\mathcal{B}$ (i.e., context) such that $\sum_{j \in \mathcal{B}} \Pi_j = I$. The ABL formula can be expressed in terms of weak values $(\Pi_j)_w = \langle \phi | \Pi_j | \psi \rangle / \langle \phi | \psi \rangle$ \cite{waegell2015contextuality},
\begin{equation}
P_{\textrm{ABL}}(\Pi_j=1 \space \; \vert \; \psi, \phi, \mathcal{B})  = \frac{  |(\Pi_{j})_w|^2  } { \sum_{k\in\mathcal{B}} | (\Pi_{k})_w|^2 }. \label{ABL}
\end{equation}
It then follows from $\sum_{k\in\mathcal{B}} (\Pi_k)_w = 1$ that $P_{\textrm{ABL}}(\Pi_j=1 \; \vert \; \psi, \phi, \mathcal{B}) = 1$ implies $(\Pi_j)_w = 1$. Furthermore, if $\mathcal{B}$ contains only two outcomes, then the converse also follows: $(\Pi_j)_w = 1$ implies $P_{\textrm{ABL}}(\Pi_j=1 \; \vert \; \psi, \phi, \mathcal{B}) = 1$. As shown in Ref.~\cite{pusey2015logical}, the ABL formula constrains any NCHVT since a projection with an ABL probability of 1 must also be assigned a value of 1 in any NCHVT.  Thus, in this case, measuring a projector weak value $(\Pi_j)_w$ of $1$ implies that any NCHVT must also assign $\Pi_j$ a value of $1$ --- and a value of $0$ to all projectors orthogonal to $\Pi_j$.

Specifying to $N$ independent neutron spins, we use $I,X,Y,Z$ to denote the independent spin components (Pauli matrices). We prepare the spins in the product state $|\psi\rangle = |{+}X\rangle^{\otimes N}$ (all $X$ eigenvalues +1), and postselect onto the product state $|\phi\rangle =  |{+}Y\rangle^{\otimes N}$. Since the predictions of products of $X$ and $Y$ by an NCHVT must be consistent with these boundary conditions, only products involving $Z$ are left undetermined (see Fig.~\ref{SquareSteps}b). The ABL rule then determines these values, as we now explain. 

For our specific case of $N>2$ spins, consider a product of any two spin operators $ZZ$, with spectral decomposition $ZZ = (+1)\Pi_{\rm even} + (-1)\Pi_{\rm odd}$ in terms of the rank-2 parity projectors,
\begin{align}\label{ZZ}
  \Pi_{\rm even} 
  &= \Pi_+\!\otimes\Pi_+ + \Pi_-\!\otimes\Pi_-, \\
  \Pi_{\rm odd} 
  &= \Pi_+\!\otimes\Pi_- + \Pi_-\!\otimes\Pi_+, \nonumber
\end{align}
with $\Pi_\pm \equiv |{\pm}Z\rangle\langle{\pm}Z| = (1 \pm Z)/2$. Given $|\psi\rangle$ and $|\phi\rangle$ defined above, $(\Pi_{\rm even})_w = 0$ and $(\Pi_{\rm odd})_w = 1$, and thus $(ZZ)_w = -1$. The ABL rule in Eq.~\eqref{ABL} then implies $ZZ = -1$ for all pairs of spins in any NCHVT, as illustrated in Fig.~\ref{SquareSteps}.

This pairwise constraint is the \emph{quantum pigeonhole effect} \cite{aharonov2016quantum}. To see this, let the spin eigenstates $\ket{\pm Z}$ correspond to two boxes in which pigeons may be placed. The projectors in Eq.~\eqref{ZZ} describe definite numbers of pigeons in each box, up to an exchange of boxes; i.e., $\Pi_{\rm even}$ denotes two pigeons in one box, while $\Pi_{\rm odd}$ denotes one pigeon in each. The pigeonhole principle states that if $N>2$ pigeons are placed in two boxes, then at least one box must contain multiple pigeons. However, the constraint $ZZ = -1$ for all pairs implies that, regardless of how many pigeons are placed in the two boxes, no two pigeons are ever in the same box!

\section{Witnessing BKS contextuality}\label{sec:witness}%
Following the pigeon analogy, all NCHVT assignments of definite numbers of pigeons to each box are forbidden
. The projectors corresponding to such forbidden assignments for $N=3$ are
\begin{align}\label{ZZZ}
  \Pi^{(3)}_0 
  &= \Pi_+\!\otimes\Pi_+\!\otimes\Pi_+ + \Pi_-\!\otimes\Pi_-\!\otimes\Pi_-, \\
  \Pi^{(3)}_1 
  &= \Pi_+\!\otimes\Pi_+\!\otimes\Pi_-  + \Pi_-\!\otimes\Pi_-\!\otimes\Pi_+ ,
  \nonumber \\
  \Pi^{(3)}_2 
   &= \Pi_+\!\otimes\Pi_-\!\otimes\Pi_+ + \Pi_-\!\otimes\Pi_+\!\otimes\Pi_-,
  \nonumber \\
  \Pi^{(3)}_3 
   &= \Pi_+\!\otimes\Pi_-\!\otimes\Pi_- +  \Pi_-\!\otimes\Pi_+\!\otimes\Pi_+ ,
   \nonumber 
\end{align}
which are the invariant eigenspaces of the first row of Fig.~\ref{SquareSteps} ($\Pi^{(3)}_0$ indicates all three pigeons in one box, while $\Pi^{(3)}_{1,2,3}$ are the permutations of two and one). Any NCHVT assigns 0 to all forbidden projectors. We call complete sets of forbidden projectors like this \emph{contextual bases}.

Crucially, such forbidden BKS value assignments manifest as anomalous projector weak values (with real part outside the range $[0,1]$) in the contextual basis of Eq.~\eqref{ZZZ} \cite{waegell2015contextuality}---classical assignments of pigeons to boxes must respect the range $[0,1]$. An anomaly indicates contradictory noncontextual value assignments to the corresponding context. Thus, the forbidden projectors constitute witness observables such that negative weak values imply confined BKS contextuality, and contradict the assignment of 0 by an NCHVT. These witnesses promote the logical contradiction of the quantum pigeonhole effect into an experimentally robust inequality. 

For the explicit example in Fig.~\ref{SquareSteps}, the weak value of $\Pi^{(3)}_0$ is
\begin{align}\label{wv3}
(\Pi_0^{(3)})_w 
&= \prod_{n=1}^3 \frac{1 + Z_w^{(n)}}{2} + \prod_{n=1}^3 \frac{1 - Z_w^{(n)}}{2} = -\frac{1}{2}, 
\end{align}
with $\ket{\psi}$ and $\bra{\phi}$ above, where each $n$ is a distinct spin, and $Z_w = \langle {+}Y|\, Z\, |{+}X\rangle / \langle {+}Y|{+}X\rangle = i$ is a purely imaginary single-spin weak value, implying $(ZZ)_w=(Z)_w(Z)_w = -1$. This example also illustrates a subtle point about the connection between anomalous weak values and contextuality. As discussed above, a projector weak value for a separable composite system with a negative real part is a witness of contextuality. However, the projector weak values for each spin are $(\Pi_\pm)_w = (1 \pm Z_w)/2 = e^{\pm i \pi/4}/\sqrt{2}$, which have positive real parts. Nevertheless, the product of three such weak values, $(\Pi_\pm^{\otimes 3})_w = (\Pi_\pm)^3_w = e^{\pm i 3\pi/4}/\sqrt{8}$, has a negative real part, enabling the contextuality witness. In this sense, the observation of a nonzero phase for a projector weak value on a single system already implies a contextuality witness on a larger composite system.


The logic for the above construction for $N=3$ generalizes to odd $N > 3$ (see the family of Wheel BKS-sets \cite{WA_Nqubits}). That is, all classical assignments of $N$ pigeons to 2 boxes are forbidden. Analogously to Fig.~\ref{SquareSteps}, the contextuality is confined to a context in the $N$-spin Wheel-set consisting of $N$ pairwise observables $ZZ$ arranged in a ring. All pairs $(ZZ)_w=-1$ as before, since each $Z_w = i$. We label the invariant eigenspace projectors corresponding to this context by defining the $N$-digit binary sequences $x^{(N)}_j$, $j \in 0 \ldots 2^{N-1}-1$, e.g., $x^{(3)}_0 = (0,0,0)$, $x^{(3)}_1 = (0,0,1)$, $x^{(3)}_2 = (0,1,0)$, $x^{(3)}_3 = (0,1,1)$. The weak values of the $2^{N-1}$ forbidden projectors (witness observables) in this contextual basis are then 
\begin{equation}\label{Proj}
(\Pi^{(N)}_j)_w = \prod^{N}_{n = 1} \frac{1 + (-1)^{x^{(N)}_{j,n}}\, Z_w^{(n)}}{2} + \prod^{N}_{n = 1} \frac{1 - (-1)^{x^{(N)}_{j,n}}\,Z_w^{(n)}}{2},
\end{equation}
where $x^{(N)}_{j,n}$ is the $n$th digit of $x^{(N)}_j$. As in Eq.~\eqref{wv3}, these projector weak values may be computed from $N$ single-spin $Z_w$ values. This great simplification enables us to construct all forbidden projectors for any number of spins by measuring single-spin $Z_w$. All projector weak values then evaluate to $(\Pi_j^{(N)})_w = \pm2^{-(N-1)/2}$, with a sign depending on the index.

Finally, we construct an \emph{unbiased contextuality witness} $C^{(N)}$, using all $2^{N-1}$ rank-2 projectors in an $N$-spin contextual basis, that aggregates the contextuality of the entire basis,
\begin{equation}\label{C}
C^{(N)} =  I^{(N)} - \sum_{j=0}^{2^{N-1}-1} s_j\, \Pi^{(N)}_j,
\end{equation}  
with $s_j =  \textrm{sign}[\text{Re}(\Pi_j^{(N)})_w]$, using the predicted value of $(\Pi_j^{(N)})_w$. Regardless of the signs $s_j$, if all $0 \leq \text{Re}(\Pi_j^{(N)})_w \leq 1$, then $\text{Re}\,C^{(N)}_w \geq 0$. Observing $\text{Re}\,C^{(N)}_w < 0$ is thus an experimental witness of confined BKS-contextuality.  This choice of the signs $s_j$ optimizes $C^{(N)}_w$ by accumulating anomalous parts of the weak values (below 0 or above 1), producing the ideal values $\text{Re}\,C^{(N)}_w = 1 - 2^{(N-1)/2}$.

\begin{figure}[t]
\centering
 \includegraphics[width=\columnwidth]{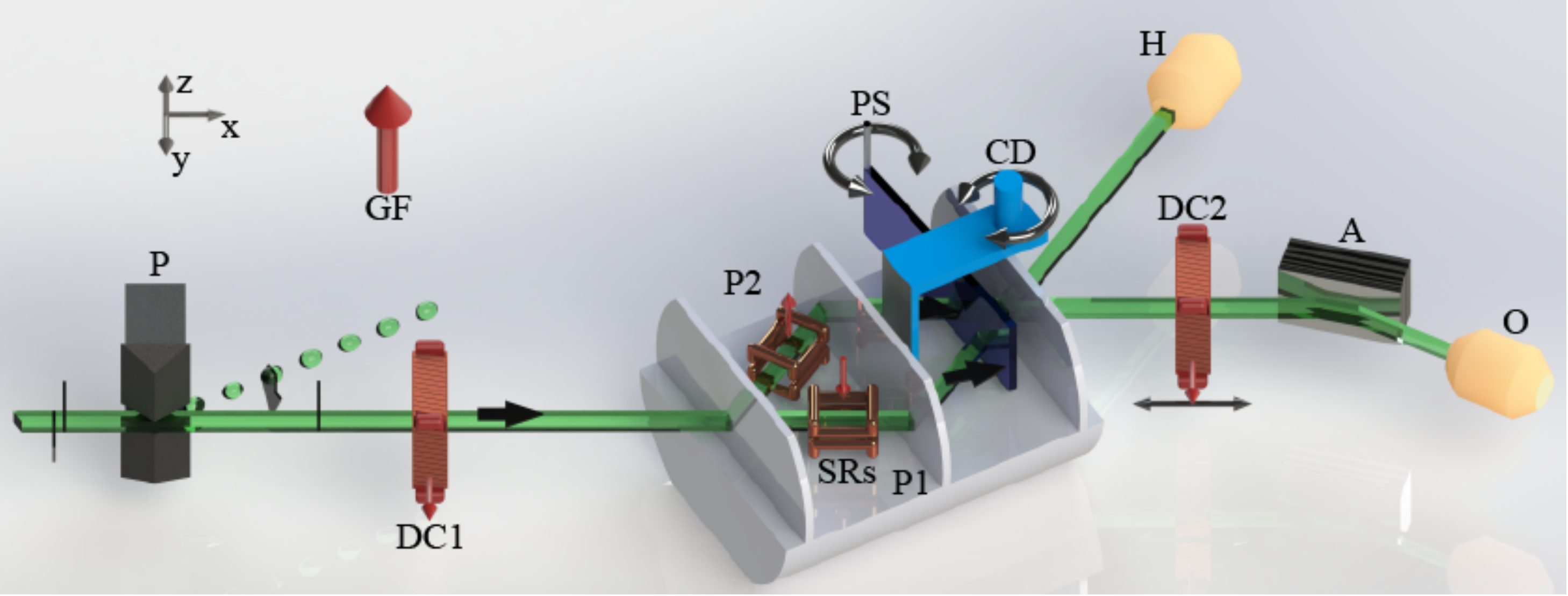}
\caption{Experimental setup. The unpolarized neutron beam passes a magnetically birefringent prism (P) that permits only spin-up neutrons to fulfill the Bragg condition for entering the interferometer. To prevent depolarization, a magnetic guide field (GF) is applied over the whole setup. A DC coil (DC1) aligns the incoming neutron spin along the positive $x$ direction. Inside the interferometer, the neutrons split into two paths (P1) and (P2), where two spin rotators (SRs) can independently rotate the neutron spin in the $xy$ plane. A cadmium slab (CD) can optionally block one of the paths. To tune the relative phase $\chi$ between the path eigenstates, a phase shifter (PS) is inserted into the interferometer. After the interferometer, a second DC coil (DC2) mounted on a translation stage, in combination with a polarizing supermirror (A), postselects a specific spin component. The neutrons are detected by $^3$He detectors (O \& H).}
\label{fig:setup}
\end{figure}

\section{Experimental procedure}\label{sec:procedure}%
In our experiment, we measure the weak value $Z_w$ of the neutron spin in the $z$-direction using an interferometer.  The neutron's path is used as a pointer to measure both the real and imaginary parts of $Z_w$.  This approach has already been successfully used to completely determine weak values of massive systems~\cite{sponar2015}. 
The experiment was conducted at the instrument S18 at the high flux research reactor of the Institute Laue-Langevin (ILL) in Grenoble, France. The experimental setup is depicted in Fig.~\ref{fig:setup}.


A perfect silicon crystal selects neutrons with a wavelength of $\lambda_0=1.91$~\AA\, ($\lambda/ \lambda_0\sim 0.02$) by Bragg reflection from a white neutron beam \cite{sponar2015}. Between the monochromator and the interferometer crystal, two magnetically birefringent prisms (P) split the unpolarized beam in two beams, one with the neutron spin aligned parallel to the positive $z$-direction and one aligned antiparallel. Even though the angular separation is just four seconds of arc (exaggerated in Fig. \ref{fig:setup}), only the beam with spin up component fulfills Bragg's condition at the interferometer's first plate. The degree of polarization is above 99\% with the neutron spin state given by $\ket{+Z}$. The other beam passes through unaffected and does not further contribute to the experiment. 

\begin{figure*}[t]
\centering
 \includegraphics[width=0.75\textwidth]{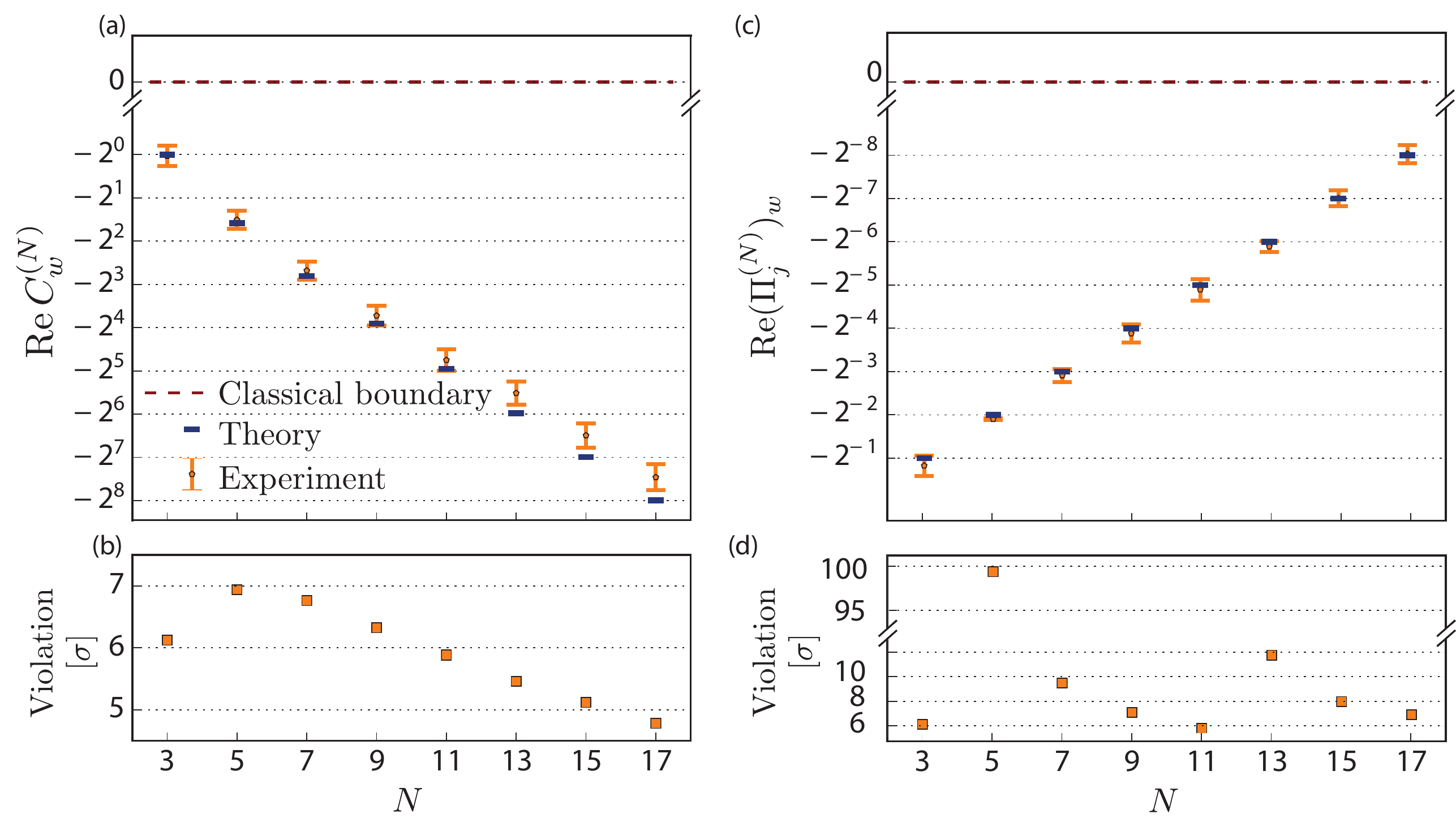}
\caption{Experimental results witnessing confined contextuality for $N$-spin Wheel BKS-sets, for $N=3,5,7,\ldots,17$. While noncontextual hidden variable predictions for the witness observables lie above the classical bound of 0 (red, dashed), quantum predictions (blue, solid) and experimental data (orange, error bars showing one standard deviation) violate this bound. (a) Unbiased witnesses $C^{(N)}_w$ given in Eq.~\eqref{C} that test an entire context, vs. $N$. (c) Exemplary set of projector witnesses $(\Pi_j^{(N)})_w$ from within a context, as given in Eq.~\eqref{Proj}, with specific indices $j$ for each $N$ given in the text. (b,d) Violation of classical boundary in units of statistical standard deviations $\sigma$, corresponding to (a) and (c), respectively. }
\label{WeakValues}
\end{figure*}

A DC coil (DC1) in front of the interferometer generates a constant magnetic field $B_y$ in $y$-direction. After entering the coil, the neutron experiences a non-adiabatic field change and its spin starts to precess around $B_y$. If the magnetic field magnitude is adjusted accordingly, the neutron spin will turn by exactly $\pi/2$ in the coil. This changes the initial spin state from $\ket{+Z}$ to $\ket{+X}$, completing the spin preselection.

At the first interferometer plate, the beam is coherently split by amplitude division. In each path, (P1) and (P2), a spin rotator (SR)---small coil in a Helmholtz configuration---produces a weak magnetic field in the $\pm z$ direction. To prevent thermal stress on the interferometer, the coils are water cooled. The weak magnetic fields lead to path-dependent spin rotations around the field axis, causing (weak) entanglement between the spin and path degrees of freedom of each neutron. For all measurements, the angle of rotation was set to $\alpha=15^\circ$. The infidelity $\mathcal{I}=\sin^2\alpha$ between the partial path states corresponding to $z = \pm 1$ quantifies the measurement strength. Our weak measurement has infidelity $\mathcal{I} = 0.067$ \cite{sponar2015}, compared to a strong measurement with $\alpha = 90^\circ$ and $\mathcal{I}=1$. 

Between the second and final interferometer plate, a sapphire phase shifter (PS) is inserted.  A phase shifter in combination with a Cadmium beam block (CD) mounted on a rotational stage provides full control over the neutron's path for the pointer readout. The phase shifter can change the path state in the equatorial plane of the Bloch sphere, while the beam block permits access to the path eigenstates at the poles.

At the final interferometer plate, the two paths are recombined. A second DC coil (DC2), in combination with a CoTi supermirror array \cite{mezei1976,mezei1977} (A), enables arbitrary spin-state postselection. 
The neutrons are detected by $^3$He counter tubes (O) and (H). Of the two outgoing ports, only (O) is analyzed to postselect the spin state $\bra{+Y}$. 

All measurements were performed using an IN/OUT method. For each fixed phase shifter position, the intensity is recorded with the spin-path coupling field turned on (IN), and then the intensity is recorded with the coupling field turned off (OUT).  Background intensities are also recorded in order to calibrate the counters.  This method permits the spin-independent relative phase $\chi$ to be determined for the path postselection state $(\Ket{P1} + e^{i\chi}\Ket{P2})/\sqrt{2}$. After curve-fitting an intensity scan over $\chi$ on the Bloch sphere equator (see Appendix \ref{B:Experiment}), the intensities for the $y$-eigenstates at points $\chi=\pi/2,\,3\pi/2$ are identified by inserting the phase values obtained from the OUT measurements into the IN measurement fit functions. These intensities determine the real part of the spin weak value $\text{Re}\, Z_w$ (Eq.~(19) of \cite{sponar2015}). This method also maximally reduces the influences of phase drift in the interferogram (due to unavoidable instability of the apparatus). To determine the imaginary part $\text{Im}\, Z_w$, it is also necessary to postselect neutron path eigenstates (Eq.~(20) of \cite{sponar2015}), which is accomplished by blocking one path at a time. If an intensity is recorded while path $P2$ is blocked, a postselection onto the state $\Ket{P1}$ is performed, and vice versa. For our choice of pre- and postselection, the expected weak value is $Z_w = i$. The negligible phase shift observed between the IN and OUT interferograms confirms that $\text{Re}Z_w \approx 0$. In contrast to that, the imaginary part shifts the pointer state towards the Bloch sphere poles, changing the relative path intensities. 

In the experiment the weak values $Z_w$ of 17 individual spins were determined with high precision. To extract $Z_w$ from one neutron spin data set, two $\chi$-scans were recorded, as well as two single intensities. Together with the required background measurements, a total collection time of ${\sim}10000$ seconds was needed to determine the real and imaginary part of each $Z_w$. 

\section{Results}\label{sec:results}%
The measured $Z_w$ are used to construct the pairwise anticorrelations $(ZZ)_w \approx -1$ (see Appendix \ref{B:Experiment}), and the $N$-spin witnesses in Eq.~\eqref{Proj} and \eqref{C}. 
Fig.~\ref{WeakValues}a,b shows the final results that violate the noncontextuality bound $\text{Re}\,C^{(N)}_w \geq 0$. Fig.~\ref{WeakValues}c,d shows final results that violate independent noncontextuality bounds $\text{Re}(\Pi^{(N)}_j)_w \geq 0$. The contextuality witnesses $C^{(N)}_w$ and $(\Pi^{(N)}_j)_w$ were calculated using Eqs.~\eqref{C} and \eqref{Proj}, respectively, for all odd numbers of spins from $N=3$ to 17. Note that the pair of forbidden projectors $\Pi^{(5)}_0$ and $\Pi^{(13)}_0$ have the remarkable geometric property that first order errors vanish when $Z_w = i$, explaining the small statistical standard deviation $\sigma$ observed in the experimental data. The chosen witnesses for other $N$ in Fig.~\ref{WeakValues}c,d are the projectors $\Pi^{(3)}_0$, $\Pi^{(7)}_1$, $\Pi^{(9)}_3$, $\Pi^{(11)}_7$, $\Pi^{(15)}_1$, and $\Pi^{(17)}_3$. The data for $N=5$ is most statistically significant, with $\text{Re}\,C^{(5)}_w = -2.85 \pm 0.41$ violating the bound of 0 by ${\sim}7\sigma$, and $\text{Re}(\Pi^{(5)}_0)_w = -0.2508 \pm 0.0025$ by ${\sim}99\sigma$. 

\section{Concluding remarks}\label{sec:conclusion}%
We have experimentally shown the confinement of contextuality within a BKS-set of observables to a particular measurement context, using modern techniques in neutron interferometry to measure weak-valued contextuality witnesses. Using $N$-spin Wheel BKS-sets \cite{WA_Nqubits}, we have reduced the problem of witnessing contextuality to weakly measuring a particular context, consisting of neighboring pairs of observables $ZZ$ arranged in a ring, with the remaining observables in the BKS-set fixed by a particular pre- and postselection. It follows that $(ZZ)_w=-1$ for all such pairs, implying anticorrelation that violates the classical pigeonhole principle \cite{aharonov2016quantum}. Moreover, the weak values of the invariant subspace projectors $(\Pi_j^{(N)})_w$ of this context contain anomalies, witnessing the failure of classical value assignments. Our unbiased contextuality witness $C^{(N)}_w$ uses all such projector weak values within the context to witness the same failure. 

Unlike the implicit global contradictions inherent to existing BKS experiments \cite{kirchmair2009state,bartosik2009,d2013experimental}, our method confines the apparent contradiction to a particular context, where its physical consequences may be explicitly revealed through weak measurements. Notably, unlike existing approaches to demonstrating BKS-contextuality \cite{pusey2014anomalous}, our witness does not require entangled preparations or measurements, or indeed any interaction between the different spins at all. The entangled measurement contexts that would normally be required have values that are forced by the pre- and postselection according to the geometry of the BKS-set itself, so they need not be measured. In this way, confining the contextuality serves to simplify its experimental observation.   
Such a simplification not only raises interesting foundational questions \cite{aharonov2016quantum}, but may also suggest future quantum information processing applications \cite{howard2014contextuality,abbott2012strong}.



\begin{acknowledgments}
 This research was supported (in part) by the Fetzer-Franklin Fund of the John E. Fetzer Memorial Trust and the Austrian Science Fund (FWF): Projects No. P25795-N20 and No. P24973-N20. 
T.D., H.G., D.E., T.J., S.S., and Y.H. performed the experiment and analyzed the data. T.D., M.W. and J.D. performed the error analysis and generated figures. T.D., M.W., and J.D. co-wrote the manuscript. M.W. and J.T. developed the original theory.
\end{acknowledgments}

\bibliography{Vienna_Pigeonhole_Biblio}

\appendix
\section{Theory}\label{A:Theory}
 \begin{figure}[t!]
 \centering
 \subfloat[][5-spin BKS Wheel]{
  \includegraphics[width=0.64\columnwidth]{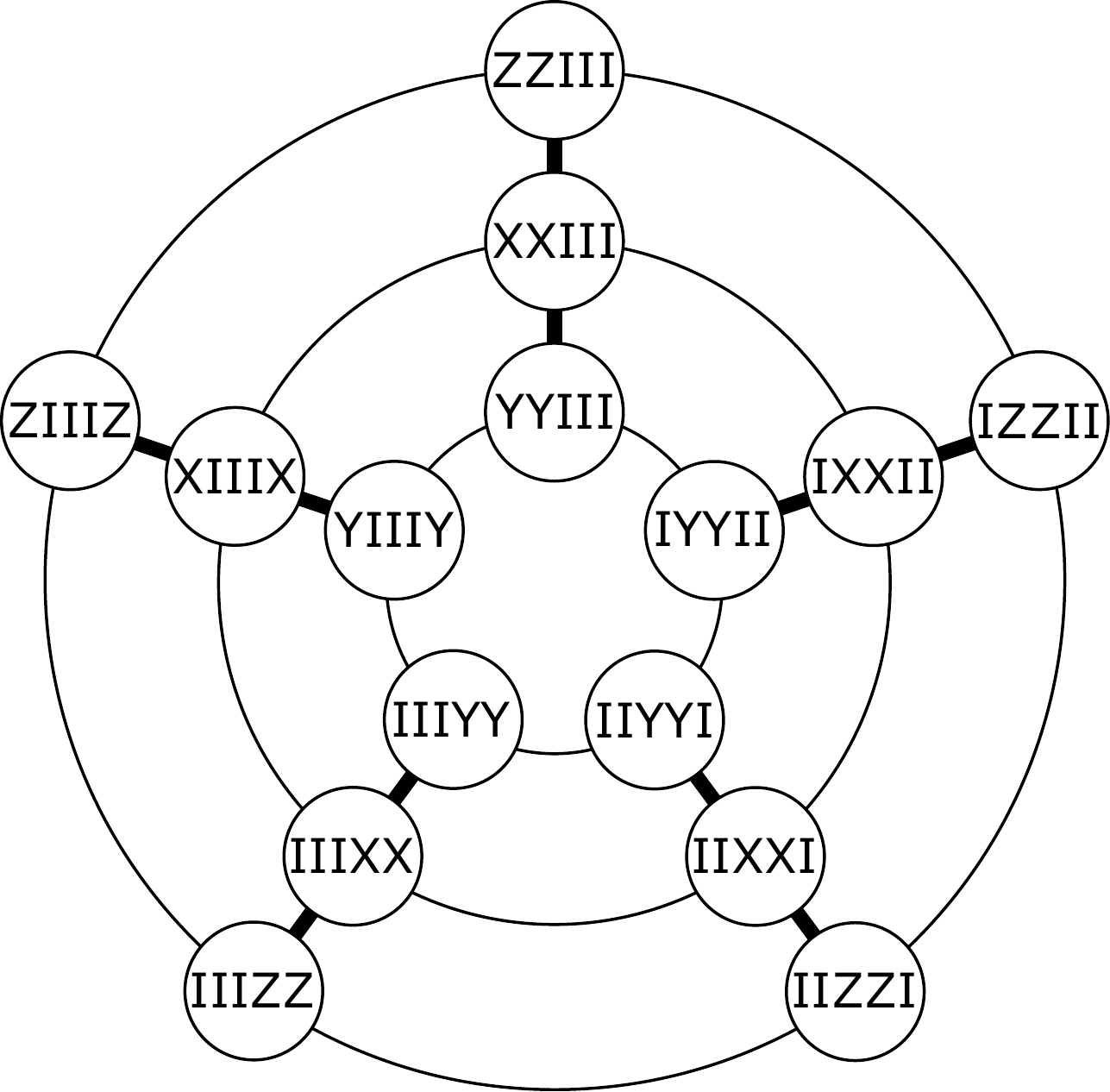}
 \label{Wheel5_0}}
 \qquad
  \subfloat[][Pre- / postselection]{
  \includegraphics[width=0.64\columnwidth]{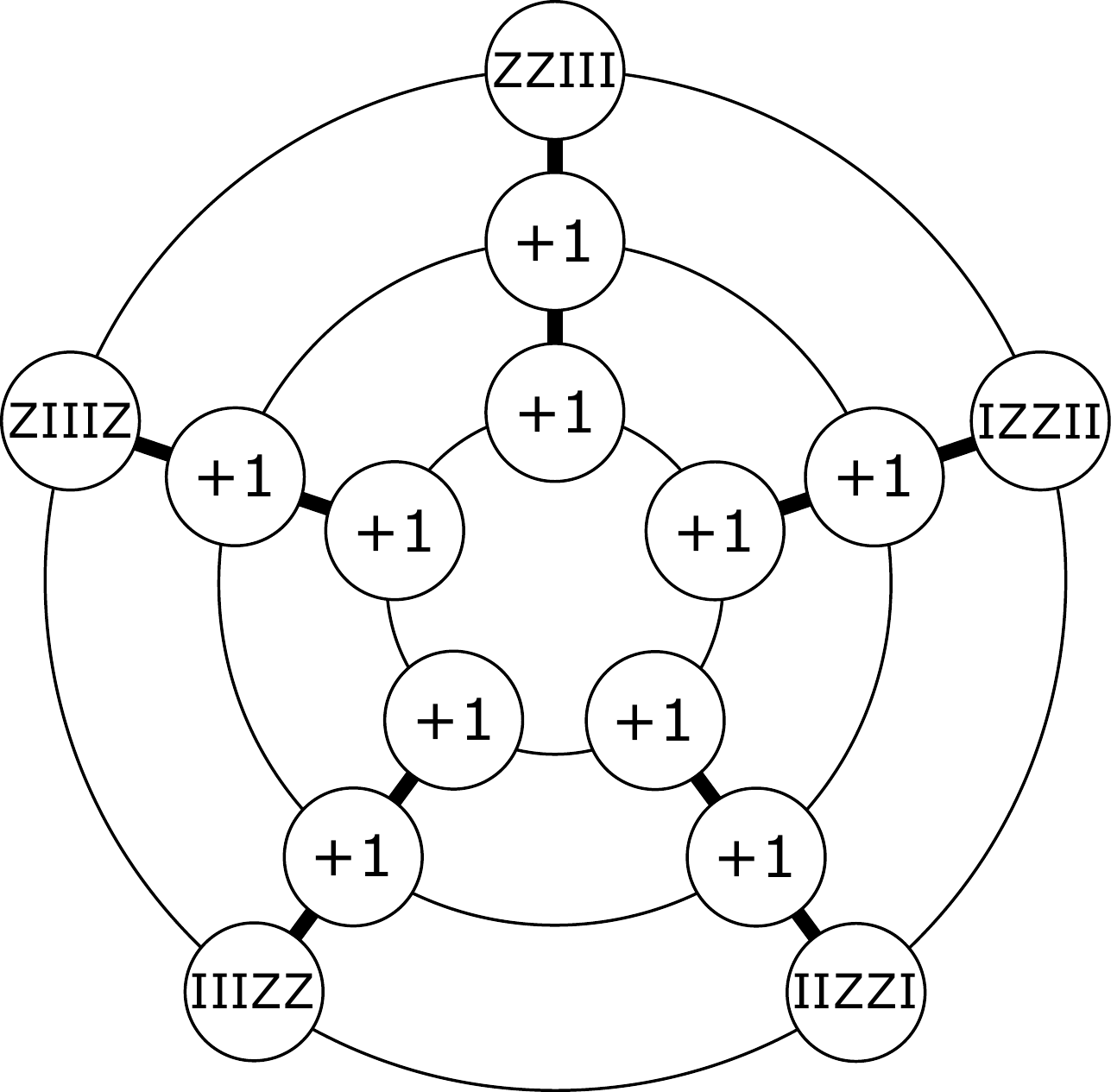}
 \label{Wheel5_1}}
 \qquad
  \subfloat[][Paradoxical Values]{
  \includegraphics[width=0.64\columnwidth]{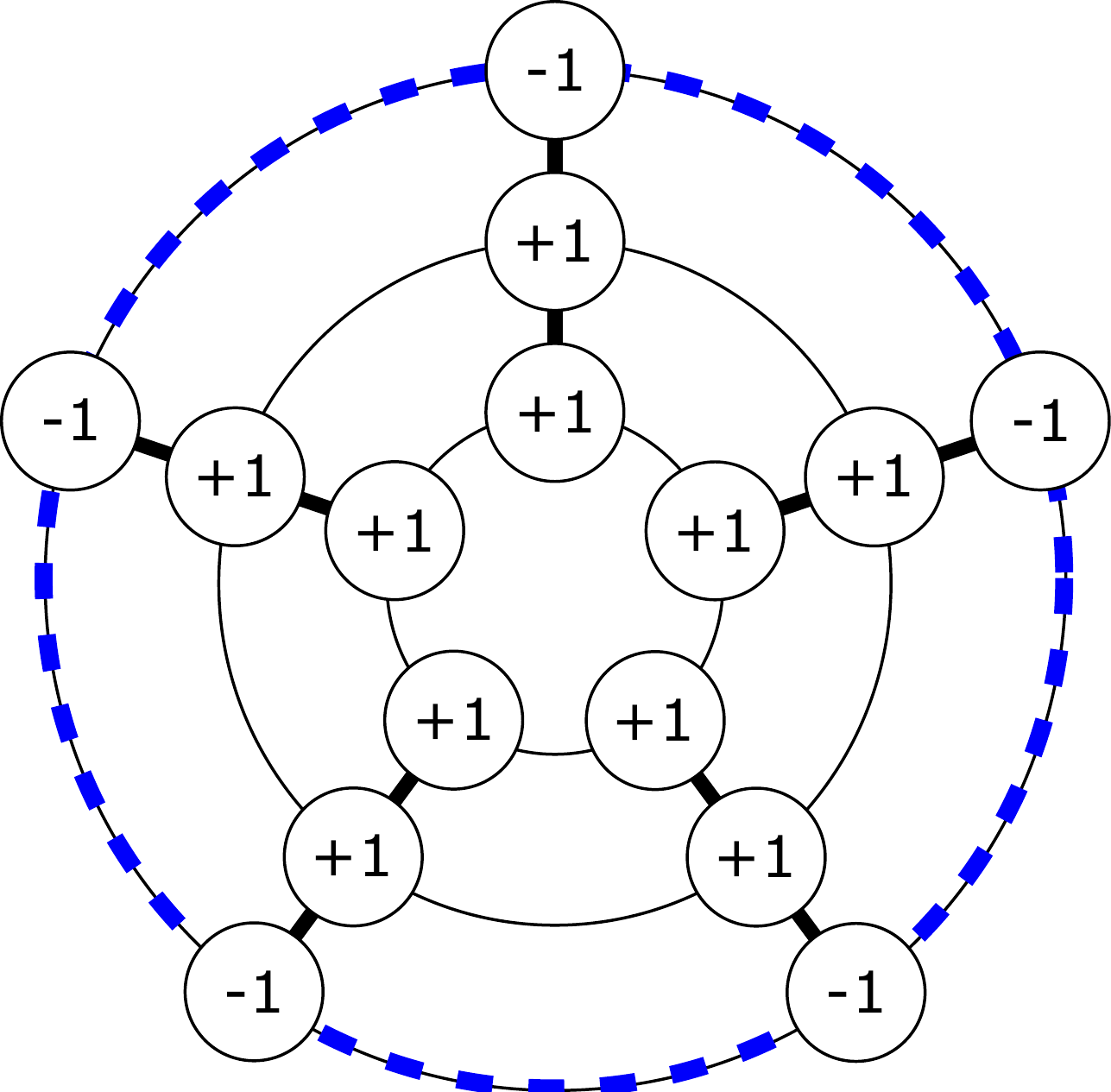}
 \label{Wheel5_2}}
 \caption{The 5-spin Wheel: (a) The product of the five observables in each ring is $+I$ (thin line), and the product of the observables in each spoke is $-I$ (thick line).  (b) A particular preparation and postselection fixes the values of two rings. (c) In any noncontextual hidden variable theory, the remaining values must be $-1$, which confines the BKS contradiction to the outer ring (blue dashed circle). This also demonstrates the quantum pigeonhole paradox: all pairs in the row appear anticorrelated, which violates the classical pigeonhole principle.}\label{Wheel5}
 \end{figure}
The family of $N$-spin Wheel KS sets \cite{WA_Nqubits} prove the BKS theorem \cite{Bell2,KS} for all odd $N \geq 3$, with the 3-spin Wheel presented as the 3-spin Square \cite{WA_3qubits} in the main text for compactness, and the 5-spin Wheel shown in Fig. \ref{Wheel5}.  Each Wheel set contains three rings composed of the $N$ pairwise Pauli observables $ZZ$, $XX$, and $YY$ respectively, of neighboring pairs in a ring of $N$ spins.  Each Wheel also contains $N$ `spokes,' which contain the three observables $ZZ$, $XX$, and $YY$ for a particular neighboring pair in the ring.  Each ring and spoke contains a set of mutually commuting observables that define a joint measurement basis.  The product of the observables in each ring (spoke) is $+I$ ($-I$), and thus quantum mechanics predicts that the product of the measurement outcomes for the observables in each ring (spoke) is $+1$ ($-1$).

A noncontextual hidden variable theory that assigns an eigenvalue prediction $ \pm 1$ to each of the $3N$ observables must violate at least one of these product predictions, which proves the BKS theorem.  To see this, consider the overall product of the predicted eigenvalues along each ring and along each spoke.  According to the quantum predictions, this product must be $-1$, since there are odd number of spokes.  However, for any noncontextual value assignment this product is $+1$, since each observable appears in one ring and one spoke, and thus all of the $3N$ eigenvalue predictions are squared in the overall product.

Preparation of $|+X\rangle^{\otimes N}$ and postselection of $|+Y\rangle^{\otimes N}$ fixes all of the pairwise $XX$ and $YY$ observables in the $N$-spin Wheel to have eigenvalue assignment $+1$ in a noncontextual hidden variable theory, and the Aharonov-Bergmann-Lebowitz rule shows that all of the $ZZ$ observables have assignment $-1$.  This results in a violation of the classical pigeonhole principle as well an apparent violation of the quantum prediction for the product in the $N$-spin pairwise $ZZ$-ring context.   The joint eigenspaces of this context are the projectors $\Pi^{(N)}_j$ of the main text, which together form the composite observable $C^{(N)}$ for the ring.  These observables witness contextuality when weakly measuring them reveals negative weak values.
\begin{figure}[h]
\centering
 \includegraphics[width=1\columnwidth]{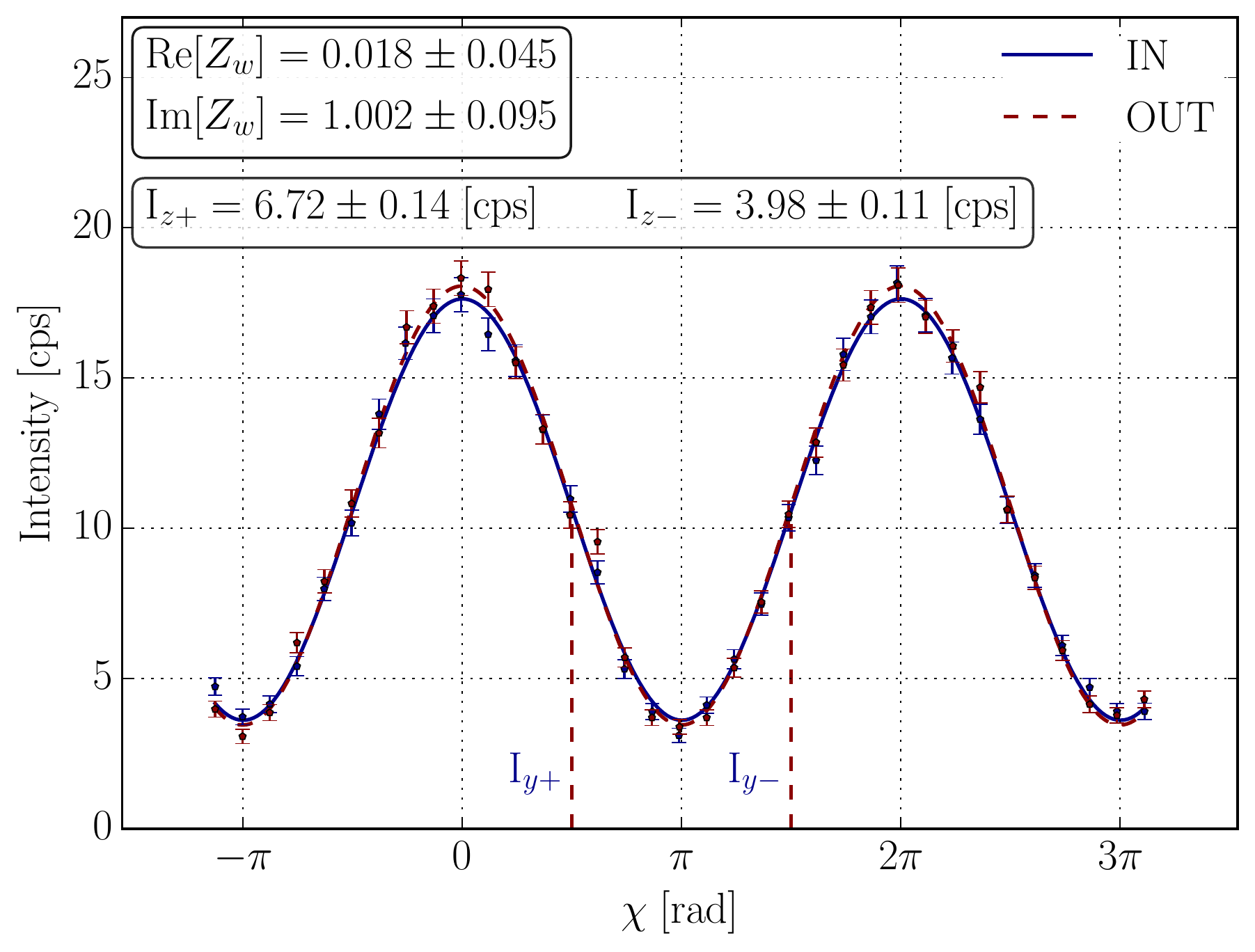} \caption{Measured interferogram for one data set: Since the weak value's real part is zero, no phase shift is seen between the IN and the OUT curve. $\text{I}_{z\pm}$ are obtained by two single intensity measurements. Background has already been subtracted.}\label{fig:set16}
\end{figure}
\section{Experiment}\label{B:Experiment}
\begin{figure}[t]
     \centering
     \begin{tabular}{|c|r|r|}\hline
       {\bf Set} &\boldmath{$\operatorname{Re}\left\lbrack Z_{w} \right\rbrack$}&\boldmath{$\operatorname{Im}\left\lbrack Z_{w} \right\rbrack$}\\\hline
       {\bf \#~1}           &$-0.024\pm0.044$&$0.970\pm0.094$\\\hline
       {\bf \#~2}           &$-0.005\pm0.044$&$1.050\pm0.098$\\\hline
       {\bf \#~3}           &$0.018\pm0.045$&$1.002\pm0.095$\\\hline
       {\bf \#~4}           &$-0.103\pm0.045$&$0.925\pm0.092$\\\hline
       {\bf \#~5}           &$-0.032\pm0.044$&$0.979\pm0.094$\\\hline
       {\bf \#~6}           &$-0.097\pm0.045$&$1.024\pm0.096$\\\hline
       {\bf \#~7}           &$0.002\pm0.049$ & $0.912\pm0.099$\\\hline
       {\bf \#~8}           &$-0.041\pm0.049$&$0.985\pm0.102$\\\hline
       {\bf \#~9}           &$0.101\pm0.051$& $0.920\pm0.099$\\\hline
       {\bf \#~10}           &$-0.020\pm0.050$&$0.931\pm0.099$\\\hline
       {\bf \#~11}           &$0.022\pm0.050$&$1.037\pm0.104$\\\hline
       {\bf \#~12}           &$-0.070\pm0.049$&$0.874\pm0.095$\\\hline
       {\bf \#~13}           &$0.011\pm0.049$&$0.790\pm0.092$\\\hline
       {\bf \#~14}           &$0.062\pm0.050$&$0.910\pm0.096$\\\hline
       {\bf \#~15}           &$-0.084\pm0.051$&$1.039\pm0.105$\\\hline
       {\bf \#~16}           &$-0.121\pm0.052$&$0.973\pm0.103$\\\hline
       {\bf \#~17}           &$-0.079\pm0.050$&$1.003\pm0.103$\\\hline
     \end{tabular}
      \caption{Experimentally determined weak values for 17 different data sets.  We used the first $N$ spin weak values shown here for our analysis of $N$-qubit contextuality witnesses in the main text.}\label{Z}
\end{figure}
\begin{figure}[t]
            \begin{tabular}{|c|r|r|}\hline
       {\bf Sets} &\boldmath{$\operatorname{Re}\left\lbrack (ZZ)_{w} \right\rbrack$}&\boldmath{$\operatorname{Im}\left\lbrack (ZZ)_{w} \right\rbrack$}\\\hline
{\bf \# 1,2} & $-1.020\pm0.137$ & $-0.030\pm0.063$ \\\hline
{\bf \# 2,3} & $-1.050\pm0.140$ & $0.014\pm0.065$ \\\hline
{\bf \# 3,4} & $-0.929\pm0.127$ & $-0.087\pm0.062$ \\\hline
{\bf \# 4,5} & $-0.902\pm0.125$ & $-0.130\pm0.061$ \\\hline
{\bf \# 5,6} & $-0.999\pm0.135$ & $-0.128\pm0.064$ \\\hline
{\bf \# 6,7} & $-0.934\pm0.134$ & $-0.086\pm0.066$ \\\hline
{\bf \# 7,8} & $-0.898\pm0.135$ & $-0.035\pm0.066$ \\\hline
{\bf \# 8,9} & $-0.910\pm0.135$ & $0.062\pm0.068$ \\\hline
{\bf \# 9,10} & $-0.859\pm0.130$ & $0.076\pm0.067$ \\\hline
{\bf \# 10,11} & $-0.966\pm0.141$ & $-0.000\pm0.070$ \\\hline
{\bf \# 11,12} & $-0.908\pm0.134$ & $-0.053\pm0.067$ \\\hline
{\bf \# 12,13} & $-0.691\pm0.110$ & $-0.046\pm0.058$ \\\hline
{\bf \# 13,14} & $-0.718\pm0.113$ & $0.059\pm0.060$ \\\hline
{\bf \# 14,15} & $-0.951\pm0.138$ & $-0.012\pm0.070$ \\\hline
{\bf \# 15,16} & $-1.000\pm0.148$ & $-0.207\pm0.075$ \\\hline
{\bf \# 16,17} & $-0.966\pm0.144$ & $-0.198\pm0.073$ \\\hline
{\bf \# 1,3} & $-0.972\pm0.132$ & $-0.007\pm0.062$ \\\hline
{\bf \# 1,5} & $-0.949\pm0.130$ & $-0.055\pm0.061$ \\\hline
{\bf \# 1,7} & $-0.885\pm0.129$ & $-0.020\pm0.062$ \\\hline
{\bf \# 1,9} & $-0.895\pm0.129$ & $0.076\pm0.065$ \\\hline
{\bf \# 1,11} & $-1.010\pm0.140$ & $-0.004\pm0.067$ \\\hline
{\bf \# 1,13} & $-0.767\pm0.116$ & $-0.008\pm0.059$ \\\hline
{\bf \# 1,15} & $-1.010\pm0.141$ & $-0.106\pm0.068$ \\\hline
{\bf \# 1,17} & $-0.971\pm0.137$ & $-0.101\pm0.066$ \\\hline
     \end{tabular}
     \caption{Experimentally determined weak values for pairwise products, showing anticorrelations between each neighboring pair in closed rings of $N$ spins, for all odd $3 \leq N \leq 17$.  The pairwise anticorrelations in these rings violate the classical pigeonhole principle.}\label{ZZC}
\end{figure}
To determine the weak value of the Pauli spin operator $Z$ the spin degree of freedom is weakly coupled to the path degree of freedom~\cite{sponar2015}. As described in the main body of the paper the weak value's real part is then inferred from an interference fringe, while two single intensity measurements are necessary to determine the weak value's imaginary part.  To determine $Z_w$ three interference fringes are recorded:
\begin{enumerate}
\item The OUT curve with no interaction, to evaluate the phase of the empty interferogram.
\item The IN curve with a path-dependent spin rotation of $\alpha=15^\circ$ and a (weak) interaction strength of $\sin^2(\alpha) = 0.067$ in each of the interferometer's arms, which yields $\text{I}_{y\pm}$.
\item One interference fringe with orthogonal preparation and postselection spin states, which is then subtracted from the IN/OUT curve as an effective background.
\end{enumerate}

Additionally two single intensities with one or the other beam blocked are recorded ($\text{I}_{z\pm}$), and again background measurements with orthogonal preparation and postselection states are performed and subtracted from the signal.  Figure~\ref{fig:set16} shows a typical IN and OUT curve of one experimental run.  The data for a complete phase-shifter scan of $\chi$ is fit to a sine function, which allows us to determine the intensity at the correct values of $\chi$, with statistical uncertainty. The measurement procedure was repeated until altogether 17 different data sets were recorded. For each data set the real and imaginary part of the the Pauli spin operator's weak value is extracted. The results are listed in Fig. \ref{Z}, and the relevant pairwise correlations are listed in Fig. \ref{ZZC}.  It is also noteworthy that the errors of sets 1 to 6 are smaller than the others due to a change in reactor power. While the first six interferograms were recorded at a power of $\sim58$~MW, for the last eleven a power was $\sim43$~MW. The increase in reactor power leads to an increase in neutron flux, and the higher count rate offers better statistics and reduces the uncertainty of the recorded values.

\end{document}